# Electric-Field Control of Quantum Tunneling Regimes in Focused He-Ion-Beam-Irradiated Oxide Interfaces

Yu Chen,[†] Maria D'Antuono,[‡,†] Robin Hutt,[¶] César Magén,[§] Edward Goldobin,[∥] Dieter Koelle,[∥] Reinhold Kleiner,[∥] Marco Salluzzo,[†] and Daniela Stornaiuolo[*,‡,†]

[†]CNR-SPIN, Complesso di Monte Sant'Angelo–Via Cinthia, I-80126 Napoli, Italy
[‡]University of Naples Federico II, I-80126 Napoli, Italy
[¶]Physikalisches Institut, Center for Quantum Science (CQ) and LISA$^+$, University of Tübingen, Auf der Morgenstelle 14, 72076 Tübingen, Germany
[§]Instituto de Nanociencia y Materiales de Aragón (INMA), CSIC–Universidad de Zaragoza, 50009 Zaragoza, Spain
[∥]Physikalisches Institut, University of Tübingen, Auf der Morgenstelle 14, 72076 Tübingen, Germany

E-mail: daniela.stornaiuolo@unina.it

## Abstract

Helium focused ion beam irradiation enables the fabrication of tunnel field-effect transistors based on two-dimensional electron systems (2DESs) at an oxide interface. High-resolution scanning transmission electron microscopy and strain mapping reveal localized lattice deformation confined to the irradiated regions, which act as nanoscale potential barriers. The barrier profile can be continuously tuned by electrostatic backgating at low temperature without degrading the electronic properties of the 2DES electrodes. Transport measurements demonstrate controlled access to thermionic emission, direct tunneling, and Fowler–Nordheim tunneling within a single device architecture. These results establish He–FIB irradiation as a powerful tool for nanoscale functional engineering of complex-oxide interfaces and provide a platform for exploring gate-tunable quantum tunneling phenomena.

## Introduction

After several years of intensive research into the fundamental properties of oxide two-dimensional electron systems (2DESs), particularly at the LaAlO$_3$/SrTiO$_3$ (LAO/STO) interface, the potential for exploiting these systems in a wide range of electronic applications has become increasingly evident.[1] This perspective has motivated the development of various oxide-based device architectures. Among them, Josephson junctions and planar tunnel field-effect transistors (TFETs) are especially attractive, both for prospective lowpower electronic applications and for the insight they provide into charge and Cooper-pair transport mechanisms. The realization of such planar devices, however, remains particularly challenging because the 2DES is buried beneath the LAO overlayer at the interface with STO. Only a limited number of Josephson junction implementations have been reported to date. These include weak links with channel lengths comparable to the superconducting coherence length,[2,3] as well as junctions defined by locally depleting the 2DES using top gates to form a tunable barrier.[4] LAO/STO-based TFETs incorporating adjustable tunnel barriers were realized using electron-beam lithography combined with reactive ion etching to define a 100nm gap between source and drain electrodes, and a lateral gate electrode for barrier tuning.[5] In these devices, charge transport was dominated by Schottky barriers with the electron gas at both sides of the gap. Owing to the relatively large barrier width, no signatures of direct tunneling were observed down to 1.2K; instead, transport was governed by thermionic emission. An alternative strategy to realize TFET relied on conducting atomic force microscopy (c-AFM) writing,[6] in which source and drain nanowires separated by a 16nm gap were written directly at the LAO/STO interface using a biased AFM tip. A third nanowire, oriented perpendicular to the transport channel, served as a lateral gate to modulate the barrier region. In these devices, transport between room temperature and 150K was dominated by thermally activated conduction, while in the temperature range from 150K down to 15K (the lowest temperature investigated), Fowler–Nordheim (F–N) field emission became the dominant transport mechanism. Electrostatic modulation of the barrier height via an applied gate voltage was also demonstrated.

In this work, we present an approach for realizing tunnel field-effect transistors based on local helium focused ion beam (He-FIB) irradiation. In recent years, He-FIB has been successfully employed to fabricate complex oxide-based devices, including nanoscale Josephson junctions and superconducting quantum interference devices based on high-temperature superconductors.[7–11] Here, we demonstrate its first application to an oxide 2DES. We show that He-FIB irradiation enables the formation of controllable nanoscale tunnel barriers with widths down to a few nanometers. The resulting TFET devices can be tuned across three distinct transport regimes: thermally activated transport at high temperatures, and both direct tunneling and Fowler–Nordheim tunneling at low temperatures. The crossover between the latter two regimes can be controlled via back-gate voltage. Transport measurements are complemented by high-resolution scanning transmission electron microscopy (STEM), which confirms the formation of a deformed region in the irradiated area that locally suppresses the conductivity of the 2DES.

**Results and Discussion**

LAO/STO (001) heterostructures are grown by reflection high-energy electron diffraction (RHEED)-assisted pulsed laser deposition. Subsequently, thin lines are irradiated using a He-FIB (see Supporting Information). Direct structural insight into the effects of He irradiation on the LAO/STO interface is obtained by atomic resolution STEM. Fig. 1(a) and (c) display images of samples irradiated with doses of 320ions/nm and 160ions/nm, respectively. At the higher dose [Fig. 1(a)], a pronounced fracture is observed along the irradiation line, extending several tens of nanometers into the STO substrate. In contrast, no obvious structural discontinuities are visible in the lower-dose sample [Fig. 1(c)]. A more quantitative understanding of the irradiation-induced modifications is obtained through strain analysis. Fig. 1(b) and (d) represent the in-plane lattice strain maps ($\epsilon_{xx}$) of the irradiated areas, where the regions of the undeformed substrate, far from the pattern, are used as a reference. Horizontal profiles of the in-plane strain, averaged vertically to reduce the noise (indicated by the dashed white rectangles), have been extracted from the STO region across the irradiation line, marked with a white arrow. These profiles, shown in Fig. 1(e) reveal strain levels between 1% and 2.5%, extending over several tens of nanometers: about 60nm and 40nm for irradiation doses of 320ions/nm and 160ions/nm, respectively. Previous studies investigating the influence of strain on the LAO/STO system, induced either by epitaxial mismatch[12,13] or high-energy proton irradiation,[14] have demonstrated that strain of only a few percent is sufficient to significantly suppress the conductivity of the 2DES and, in some cases, drive it insulating[12,13] via localization effects.[14] These studies motivated us to investigate the transport properties of devices in which He-FIB irradiation is employed to locally generate a potential barrier within the 2DES. With this aim, we fabricated a series of LAO/STO channels, each 100μm wide, with the geometry illustrated in Fig. 2(a). Along each channel, three narrow regions were irradiated using different doses (red areas in the figure). The contact configuration allows each irradiated region to be electrically characterized independently, as exemplified for device 2 in Fig. 2(a), while two devices share identical unirradiated channel regions acting as electrodes. This design enables a comparison of irradiation effects under identical growth and processing conditions and excludes artifacts arising from possible film inhomogeneities. Fig. 2(b) shows a representative current–voltage (*I–V*) characteristic measured at 5K for device A2 (Sample A) three weeks after irradiation with a dose of 50ions/nm (the delay due to sample transfer and measurement preparation). The device exhibits a pronounced bias dependence of the differential resistance at low voltages, characteristic of tunneling regime. As the bias voltage increases, the differential resistance decreases continuously and eventually saturates at approximately 1.5MΩ. The smooth crossover between the two conduction regimes takes place around $|V_{th}|$ = 0.8V, defined as the voltage-axis intercept obtained by linear extrapolation of the high-current region of the *I–V* curve (red line in Fig. 2(b)). We find that $V_{th}$ increases with increasing irradiation dose. However, subsequent measurement campaigns conducted 6 ($t_2$) and 9 months ($t_3$) after irradiation reveal a marked decrease of $V_{th}$ with time, as shown in Fig. 2(c). In particular, for device A2, $V_{th}$ is found to vanish 6 months after irradiation (see the I–V taken at $t_2$ shown in the inset of Fig. 2(b)). The temporal evolution of He-FIB–irradiated oxide devices is discussed in detail at the end of this section. Because of this time dependence, the irradiation dose alone does not provide a reliable parameter for device classification. In the following, we therefore adopt $V_{th}$ as the relevant figure of merit, as it reflects both the initial irradiation conditions and the subsequent temporal evolution of the barrier properties.

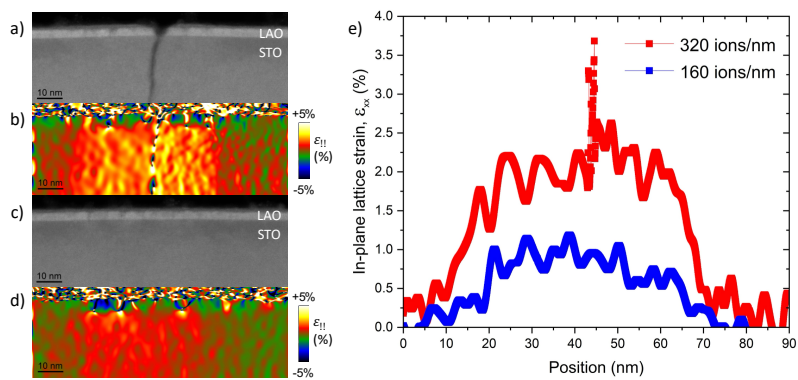

Figure 1: (a,c) Cross-sectional STEM images of LAO/STO heterostructures irradiated with He-FIB at doses of 320ions/nm (a) and 160ions/nm (c). A fracture extending into the STO substrate is observed at the higher dose, while no clear discontinuity is visible at the lower dose. (b,d). Corresponding in-plane strain maps ($\epsilon_{xx}$). The unstrained substrate lattice parameter, far from the irradiated area, is used as a reference. (e) In-plane strain profiles along the white arrows in (b) and (d), vertically integrated along the dashed rectangles.

To characterize the barrier properties in detail, we systematically analyze the I–V characteristics as a function of both back-gate voltage $V_{bg}$ and $V_{th}$ (instead of the dose value). Fig. 3 presents the I–V curves of devices B1, B2, and B3 (sample B), irradiated with different ion doses and measured at various time intervals after irradiation. The corresponding threshold voltages $V_{th}$ in their pristine state, prior to the application of any gate voltage, are 20 mV (panel a), 100 mV (panel b), and 975 mV (panel c). Notably, $V_{th}$ is progressively suppressed with increasing gate voltage and eventually vanishes in panel (a), at sufficiently large $V_{bg}$, signaling a transition to an ohmic-like transport regime (see Supporting Information for a quantitative analysis). Charge transport across the barrier can occur through several different physical mechanisms, schematically illustrated in Fig. 4(a–c), depending on the temperature and applied bias. At high temperatures, charge carriers can acquire sufficient energy to overcome the
barrier, resulting in thermionic emission (Fig. 4(a)). At lower temperatures, transport is instead dominated by quantum tunneling through the barrier. In this regime, the effective barrier profile depends on the applied bias voltage. At low bias, the barrier has a trapezoidal shape, corresponding to direct tunneling (Fig. 4(b)). As the bias increases, the barrier becomes triangular and Fowler–Nordheim (F-N) tunneling becomes the dominant transport mechanism (Fig. 4(c)). The three transport regimes give rise to distinct I–V curves. In thermionic emission regime, the current density J is described by the Richardson–Schottky (RS) model:[15]

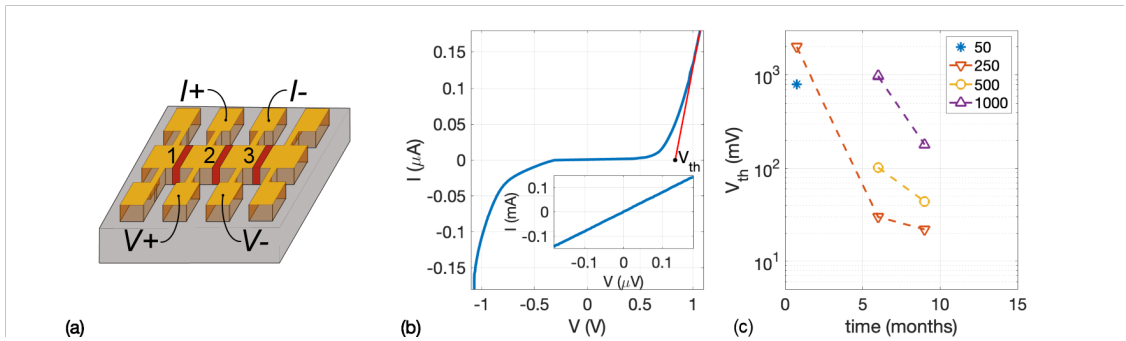

Figure 2: (a) Sketch of the devices layout and measurement geometry for device 2. (b) I-V curve of device A2 (dose: 50 ions/nm) measured at 5 K shortly after He-FIB irradiation. $V_{th}$ is defined as the voltage-axis intercept of a linear fit to the high-current region (red line). In the inset we show the I-V of the same device measured 6 months after irradiation. (c) Evolution of $V_{th}$ as a function of the time for different doses.

$$J \propto T^2 \exp\left[\frac{-(\Phi_B - \sqrt{q^3 V/4\pi\epsilon_0\epsilon_r d})}{k_B T}\right], \tag{1}$$

where $\Phi_B$ is the barrier height, $q$ is the elementary charge, $\epsilon_0$ and $\epsilon_r$ are the vacuum permittivity and relative permittivity of the barrier material, respectively, $d$ is the barrier thickness, and $k_B$ is the Boltzmann constant. At low temperatures and low bias voltage, direct tunneling through a trapezoidal barrier leads to:

$$J \propto V \exp\left[-\frac{2d\sqrt{2m\Phi_B}}{\hbar}\right], \tag{2}$$

where $m$ is the effective mass of the charge carriers. At low temperature and high bias voltage, when the barrier becomes triangular, the *J–V* characteristics are instead described by F–N tunneling:[15]

$$J \propto V^2 \exp\left[-\frac{4d\sqrt{2m\Phi_B^3}}{3\hbar q V}\right], \tag{3}$$

For large barrier height $\Phi_B$ and width $d$, F-N tunneling is expected to be the dominant observable mechanism at low temperature, while for smaller barriers, both direct and F-N tunneling may occur, depending on the applied bias. These two tunneling regimes can be readily distinguished by rewriting Eqs. (2) and (3) in logarithmic form:

$$\ln\left(\frac{J}{V^2}\right) \propto \ln\left(\frac{1}{V}\right) - \frac{2d\sqrt{2m\Phi_B}}{\hbar} \tag{4}$$

$$\ln\left(\frac{J}{V^2}\right) \propto -\left(\frac{1}{V}\right)\left(\frac{4d\sqrt{2m\Phi_B^3}}{3\hbar q}\right) \tag{5}$$

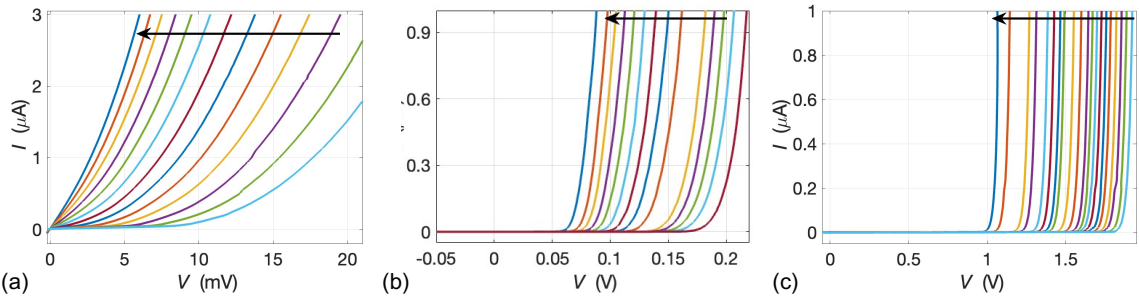

Figure 3: *I − V* curves of (a) device B1 (dose: 250ions/nm) measured at $t_3$, (b) device B2 (dose: 500ions/nm) measured at $t_2$ and (c) device B3 (dose: 1000ions/nm) measured at $t_2$ for different values of the increasing back-gate voltage (arrow) (see also Fig. S2 in the Supporting Information). All the measurements were performed at $T$ = 5 K.

Accordingly, in a plot of ln(J/V 2) versus (1/V ), direct tunneling results in a logarithmic dependence (Eq. 4) while F-N tunneling is characterized by a linear dependence with a negative slope (Eq. 5). In Fig. 4(d)–(f), we plot the data of Fig. 3 as ln(J/V 2) versus (1/V ). In all three panels, a linear behavior with a negative slope is clearly observed at high bias voltage (low 1/V ), indicating F-N tunneling. In panel (d), a logarithmic increase at low bias voltage (high 1/V ) is also visible, and a similar trend can be recognized in the low-bias region of panel (e), indicative of direct tunneling. The voltage corresponding to the inflection point, marking the crossover between the two tunneling regimes, is highlighted by a dashed gray line.

To apply Eqs. 4 and 5 to the experimental data, one of the two barrier parameters—height $\Phi_B$ or width $d$—must be independently determined. We therefore adopt the following procedure: (i) The barrier height $\Phi_B$ is first extracted from the thermionic emission regime at high temperature using Eq. 1. (ii) The value obtained from $\Phi_B$ is then used to determine From high temperature data (step (i), see Supporting Information for details) we obtain barrier heights $\Phi_B$ in the range 0.06–0.26 eV (black squares in Fig. 5), consistent with previously reported values of LAO/STO TFET.[5,6] Using these $\Phi_B$ values, we return to the low-temperature data and extract the barrier width $d$ by fitting the pristine $I - V$ characteristics in the F–N regime with Eq. 5 (step (ii)). The resulting widths range from 6.5 nm to 100 nm, in good agreement with the length scales observed in TEM analysis (Fig. 1(e)). The extracted values of $d$ are subsequently used to fit the high-bias regions of the gate-dependent curves shown in Fig. 4(d–f) using Eq. 5, allowing us to determine the field-effect-induced modulation of $\Phi_B$. Representative fits are presented in the Supporting Information. The resulting $\Phi_B$ values (triangles in Fig. 5) are in excellent agreement with those obtained from the thermionic emission analysis. For devices with small barrier widths, the same $\Phi_B$ values also provide a consistent description of the direct tunneling regime in Fig. 4(d). In contrast, for the data shown in Fig. 4(e) the direct tunneling regime becomes less distinct, while in Fig. 4(f) it is no longer observable. This systematic trend correlates with the increase in barrier width, which reaches values on the order of 100 nm for the devices in Fig. 4(f), thereby strongly suppressing the probability of direct tunneling.

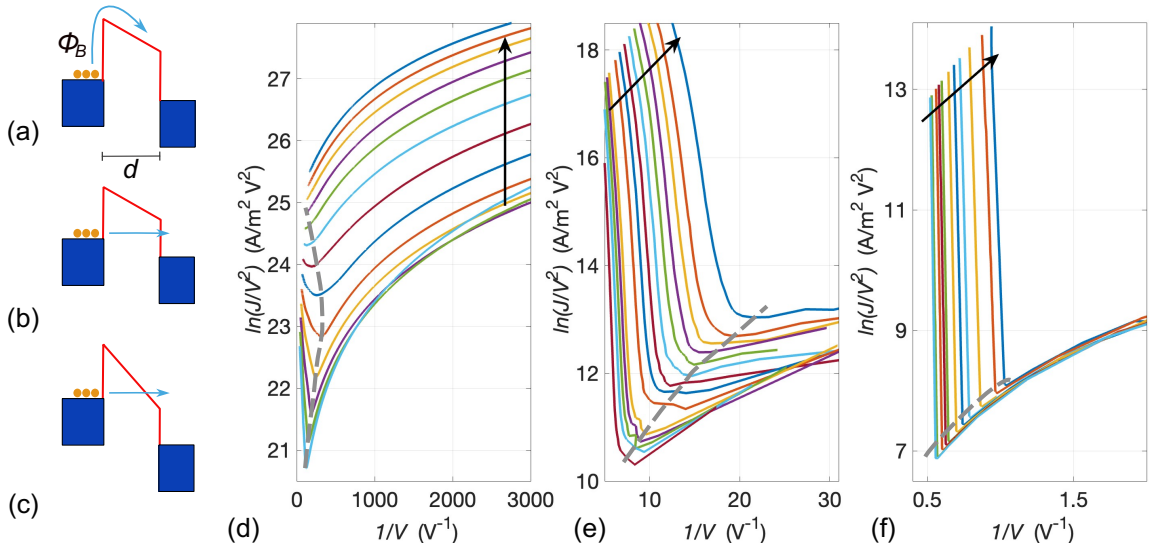

Figure 4: Sketches of three possible tunnel mechanisms: thermal activation (a), direct tunneling (b), and Fowler-Nordhaim tunneling (c). Panels (d), (e), (f) show the data of Fig. 3(a), (b), (c) respectively, plot as ln(J/V²) vs. (1/V). The arrows indicate the direction of increasing $V_{bg}$. The gray dashed line is a guide to the eye for the inflection point $V_i$, signaling the transition to F-N tunneling at high V. the barrier width $d$ from the F–N fits to the low-temperature data. (iii) Finally, extracted $d$ is used to quantify gate-induced modulation of $\Phi_B$ in both the F–N and direct tunneling regimes. The pronounced temperature dependence of the STO dielectric constant leads to substantial variations in the electric field generated by a fixed value of $V_{bg}$ at different temperatures. As a result, measurements performed at nominally identical $V_{bg}$ values are not directly comparable across temperatures. Therefore, steps (i) and (ii) were carried out in the pristine state, prior to the application of any gate voltage.

The barrier dimensions extracted for our devices are comparable to those achieved using AFM-based writing techniques (Ref.[16]) and electron-beam lithography (Ref.[5]). However, in contrast to these approaches, our devices allow access to a broader tunneling regime, encompassing both direct and F–N tunneling. We attribute this capability to the different field-effect geometry employed. The back-gate configuration used in our devices enables uniform electrostatic modulation of the channel through a robust gate dielectric layer. By comparison, the lateral-gate geometry adopted in the cited works may suffer from significant leakage currents when the dielectric separation is small, while larger separations can lead to reduced or spatially non-uniform carrier modulation.[17]

Although the devices remained stable throughout the low-temperature measurements reported here, their properties—most notably the threshold voltage $V_{th}$—were found to evolve when remeasured several months later (see Fig. 2). A temporal evolution of HeFIB–irradiated oxide-based devices properties has previously been reported by some of the present authors for Josephson junctions with barriers defined by He-FIB irradiation between YBCO electrodes.[18] In that study, stabilization times ranged from several days to several years, depending on the irradiation dose, and the effect was attributed to oxygen-vacancy migration. Structural and electrostatic relaxation effects with a comparable time-scale have also been reported in STO and STO-based heterostructures after various external perturbations, including light exposure, mechanical stress, and ion-beam irradiation.[19–21] In this case, when strain or other structural modifications, such as defect reconfiguration at the interface are introduced, the initial electronic and lattice response may be rapid, but subsequent evolution driven by defect migration, ferroelastic domain-wall rearrangement, or chemical re-equilibration can persist for days, or longer. The relevant timescales strongly depend on temperature, defect density, and sample history. Ion-irradiation studies similarly report continued vacancy diffusion and chemical equilibration well after the irradiation step, highlighting the presence of slow structural degrees of freedom.[22] Taken together, these findings indicate that STO-based systems possess multiple relaxation channels operating over different timescales, with long-time relaxation tails that can extend over months.

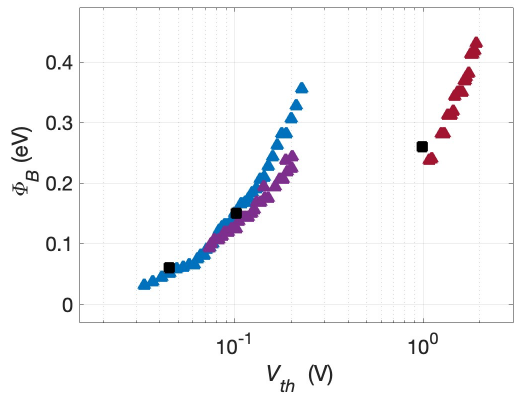

Figure 5: Barrier energy $\Phi_B$ as a function of $V_{th}$ calculated from thermal activation (black squares) and using the F-N fit of the curves shown in Fig. 4: blue, purple and dark red triangles corresponds to the data shown in panels (d), (e) and (f) respectively.

Despite these long relaxation times, the results presented in Figs. 4 and 5 demonstrate that He-FIB irradiation provides a powerful and highly localized tool for engineering transport properties in LAO/STO, enabling the formation of a nanoscale barrier which can be continuously tuned by the electric field effect, allowing controlled access to different transport regimes ranging from thermally activated conduction to direct and Fowler–Nordheim tunneling. To the best of our knowledge, this represents the first realization of a nanoscale irradiation-defined tunnel barrier in an oxide-based two-dimensional electron gas. While irradiation-induced strain effects have previously been reported in LAO/STO bilayers exposed to oxygen or proton beams,[14,23] those approaches relied on large-area irradiation that rendered most of the interface insulating, with micron-scale conducting channels preserved by resist or stencil masks. In contrast, the use of a focused helium ion beam enables the direct patterning of insulating regions with nanometer-scale lateral dimensions, offering substantially improved spatial resolution and precise control over the barrier geometry.

**Conclusions**

In this work, we demonstrate two central advances: first, the use of He–FIB irradiation to define well-controlled tunneling barriers in a LAO/STO two-dimensional electron system; and second, the ability to access and continuously tune electron transport across the full range of conduction regimes within a single tunnel field-effect transistor, spanning thermally activated transport, direct tunneling, and Fowler–Nordheim tunneling.

Beyond the specific devices investigated here, our results establish He–FIB as a powerful nanofabrication strategy for oxide electronics, enabling barrier geometries and lateral dimensions that are difficult—or in some cases unattainable—using conventional techniques such as electron-beam lithography. The combination of nanometer-scale spatial precision and controlled local defect engineering provides a versatile platform for the realization of field-effect-tunable oxide-based quantum nanodevices, including tunnel junctions, quantum point contacts, and hybrid superconducting structures. More broadly, He–FIB irradiation opens new opportunities for the controlled study of mesoscopic quantum transport in complex oxides and for the development of scalable oxide-based device architectures targeting future quantum technologies.

**Acknowledgements**

The Authors acknowledge funding from Ministero dell'Istruzione, dell'Università e della Ricerca (MIUR) for the PRIN project STIMO (Grant No. PRIN 2022TWZ9NR), from the European Union's Horizon Europe research and innovation programme for the IQARO project (grant agreement 101115190), from DFG via project GO-1106/7-1, from the Spanish Ministry of Science, Innovation and Universities (MCIN/AEI/https:// doi. org/ 10. 13039/ 50110 00110 33), through the Severo Ochoa CEX2023-001286-S and from Gobierno de Aragon project E13-23R. Authors also acknowledge the use of instrumentation as well as the technical advice provided by the National Facility ELECMI ICTS, node «Laboratorio de Microscopias Avanzadas (LMA)» at «Universidad de Zaragoza».

# Electric-Field Control of Quantum Tunneling Regimes in Focused He-Ion-Beam-Irradiated Oxide Interfaces - Supporting Information

**Samples preparation**

Epitaxial LAO films, 10 unit cells thick, are deposited on TiO$_2$-terminated STO (001) substrates by pulsed laser deposition using a KrF excimer laser (wavelength 248 nm). The thin films growth is monitored using RHEED (reflection high-energy electron diffraction).
During the deposition the substrate is kept at 680° C in oxygen partial pressure p[O2] of 1x10$^{-4}$ mbar. Following deposition, the samples are slowly cooled down to room temperature in the same oxygen pressure used for deposition.
The samples are patterned using a combination of photolithography, cold ion milling and FIB irradiation. Firstly, channels 100µm wide are realized using the cold ion milling technique described in Ref.[1] Then He-FIB is used to irradiate a thin line across them and create the barrier. He-FIB irradiation is done in a Zeiss Orion NanoFab He-ion Microscope (HIM) with 30keV He ions.[2] A beam current of 500fA is used, and the typical FIB spot-size is below 10nm for the focusing parameters used. A dwell time of 1µs is used to irradiate line patterns with a pixel spacing (pitch) of 0.25nm, which corresponds to a single line scan dose $d_{SL}$ = 12.5ions/nm. To obtain higher values $D_L$ of a line dose, a line scan was repeated $N$ times so that $D_L = Nd_{SL}$. To irradiate the rectangular areas used to perform TEM analysis the areas are irradiated line by line with the same pitch of 0.25nm between the lines. For areas the dose $D_A$ is measured in *ions/nm$^2$*. Note that since the FIB spot-size is approximately one order of magnitude larger than the pitch, a strong overlap of spots corresponding to the neighboring pixels occurs. Thus, for example, the damage caused by $D_L$ = 100ions/nm *does not* corresponds to an area dose of $D_A$ = 4 · 100ions/nm$^2$ (in 1nm$^2$ one has 4 1nm long lines) as one could naively assume. Instead, much smaller area dose causes the equivalent damage, the factor is proportional to the ratio of the spot-size and the pitch and is about 15 ± 5 in our case. Finally, the samples are measured using a variable temperature cryostat with a base temperature of 5 K. The tuning of the transport properties is carried out using electric field effect in the back gate configuration.

**STEM analysis**

Local atomic structure of LAO-STO interfaces in selected irradiation patterns is analyzed by aberration-corrected STEM on a Thermo Fisher Scientific probe-corrected Titan 60–300 microscope. The instrument is operated at 300 kV and equipped with a high-brightness
Schottky field emission gun (X-FEG), a Wien filter monochromator and a CETCOR aberration corrector for the condenser system (by CEOS). Atomic resolution Z-contrast images are obtained in high-angle annular dark field (HAADF) imaging mode. The probe has a convergence semi-angle of 24 mrad, resulting in a probe size of < 1 Å. Geometrical phase analysis (GPA) of the HAADF-STEM images is carried out to determine the relative lattice strain of the patterned areas with respect to nearby non-irradiated areas.3 The images were collected with the fast scan direction along the LAO-STO interface to minimize the effect of scan distortions and drift on the determination of the in-plane lattice strain.4

**Evolution of the devices properties with time**

In Fig. S1 we show the *I–V* curves of device B2 (*D* = 500ions/nm, panel (a)) and B3 (*D* = 1000ions/nm, panel (b)) measured 6 months after irradiation ($t_2$) and 9 months after irradiation ($t_3$).

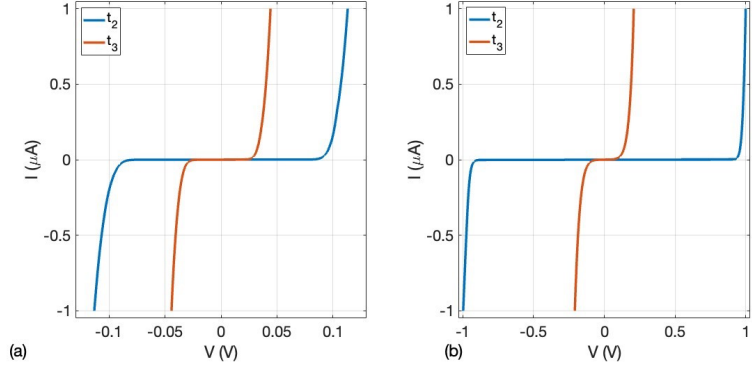

Figure S1: *I–V* curves of device B2 (a) and B3 (b) measured at different time intervals (*T* = 5K).

### $V_{th}$ evolution with the gate voltage

In Fig. S2 we plot the evolution of $V_{th}$ as a function of the applied gate voltage $V_{bg}$ for the three devices reported in Fig. 3 of the main text.

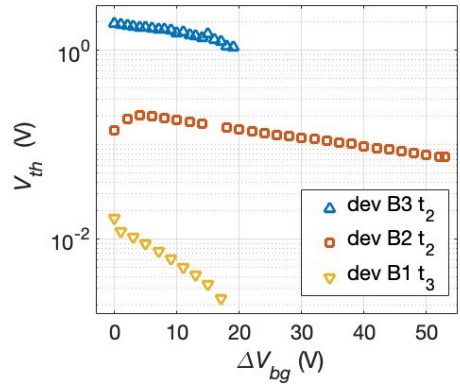

Figure S2: Vth vs. Vbg of devices B1, B2 and B3.

### Calculation of the barrier energy $\Phi_B$

According to Eq. 1, the current density at zero bias, $J_0 = J(V = 0)$, is given by $J_0 \propto T^2 \exp(-\Phi_B/k_B T)$. Therefore, the barrier height $\Phi_B$ can be directly extracted from the slope of the $\ln(J_0/T^2)$ vs. $T^{-1}$ plot. Fig. S3(a) shows the *I–V* characteristics of device B2 measured between 130 K and 150 K in the pristine state, i.e., prior to the application of any gate voltage. The same data are presented in Fig. S3(b) as $\ln(J)$ vs. $V^{1/2}$. The red lines are linear fits to the high-voltage region, from which the zero-bias intercept $J_0$ is obtained. These intercept values are then plotted as $\ln(J_0/T^2)$ vs. $T^{-1}$ in Fig. S3(c). The barrier height $\Phi_B$ is determined from the slope of the linear fit to the data in Fig. S3(c). This analysis procedure was systematically applied to all devices discussed in this work.

### Fit of the tunneling regime

After using pristine-state data to determine the barrier height $\Phi_B$ from the high-temperature thermionic emission analysis and the barrier width $d$ from F-N plots at $T = 5K$, we analyzed the gate-dependent $I$–$V$ characteristics to quantify the modulation of $\Phi_B$ induced by $V_{bg}$. Specifically, Eq. 5 was used to fit the F–N regime, keeping the previously extracted value of $d$ fixed and treating $\Phi_B$ as the only free fitting parameter. The resulting $\Phi_B$ values were subsequently inserted into Eq. 4 to model the direct tunneling regime, when observable.

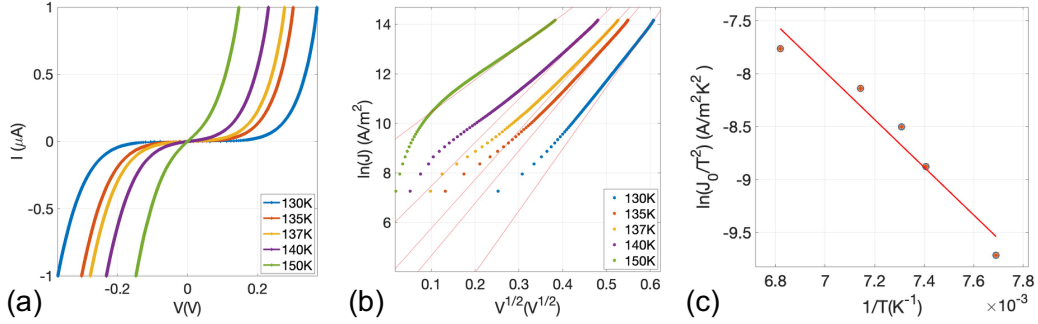

Figure S3: Fit of the high temperature regime: $I$–$V$ curves measured in pristine state for device B2 between 135 K and 150 K; (b) the same curves plot as $\ln(J)$ vs. $V^{1/2}$; the red lines are linear fit of the high V region; (c) plot of $\ln(J_0/T^2)$ versus $T^{-1}$; the red line is the linear fit, from which slope $\Phi_B$ is directly calculated.

Representative fitting curves are shown in Fig. S4. For large positive gate voltages, we observe that the nominally linear region of the $\ln(J/V^2)$ versus $1/V$ plots—characteristic of F–N tunneling—develops a double-slope behavior, with a bending at high bias. Similar deviations from ideal behavior have been reported in various material systems and have been attributed either to space-charge effects[5] or to the onset of a double-barrier transport mechanism.[6]

**Conduction mechanisms in tunnel devices**

In Table 1 we summarize the possible conduction mechanisms in tunnel devices with their temperature and voltage dependence.

| Conduction mechanism | Equation | Voltage dependence | Temperature dependence |
|---|---|---|---|
| Direct tunneling | $J \propto V e^{-\frac{2d\sqrt{2m\Phi_B}}{\hbar}}$ | $J \propto V$ | None |
| Fowler–Nordheim tunneling | $J \propto V^2 e^{-\frac{4d\sqrt{2m\Phi_B^3}}{3\hbar q V}}$ | $\ln(J/V^2) \propto 1/V$ | None |
| Thermal transport | $J \propto T^2 e^{-\frac{\Phi_B - \sqrt{q^3 V/(4\pi\epsilon_0\epsilon_r d)}}{k_B T}}$ | $\ln(J) \propto V^{1/2}$ | $\ln(J/T^2) \propto 1/T$ |

Table 1: Conduction mechanisms in tunnel devices, adapted from Ref.[7]

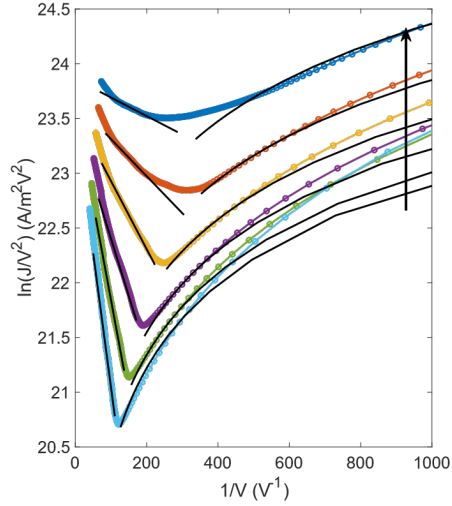

Figure S4: I–V curves of device B1 as a function of increasing gate voltage (arrow). The black lines are fits obtained using the direct and Fowler–Nordheim models. In the Fowler–Nordheim regime a double-slope behavior is observed.

**Additional data on the barrier energy**

In the low temperature regime, the transition from direct tunneling to F-N tunneling takes place when the bias voltage exceeds the barrier height (see Fig. 4(b) and (c)). Therefore, the inflection point $V_i$ in a $\ln(J/V^2)$ vs $(1/V)$ plot can give an indication of $\Phi_B$. In Fig. 4 we have indicated the voltage position of such inflection point using a dashed grey line. In Fig. S5 we report the values of $V_i$ (expressed in eV) as a function of the threshold voltage $V_{th}$, to compare them with the barrier height values obtained from the F-N and thermal regime fits. The agreement among the data is excellent.

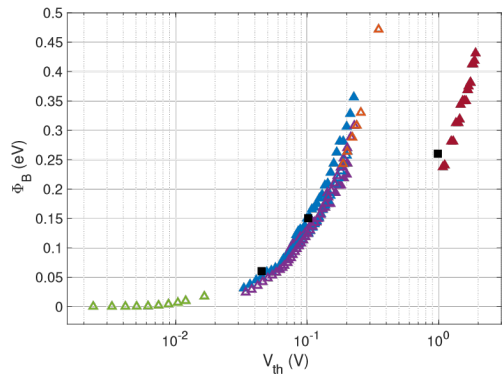

Figure S5: The data presented in Fig. 5 (solid symbols) are compared with the values of the inflection voltage $V_i$ (open symbols).